\documentclass[12pt, a4paper]{article}

\usepackage{amsmath,amssymb,amsfonts}
\usepackage{graphicx}
\usepackage[colorlinks,hypertexnames=false]{hyperref}
\usepackage{cite}

\begin{document}

\title{Dynamical friction in rotating ultralight dark matter galactic cores}
\author{V.M.~Gorkavenko${}^{1,}$\thanks{Corresponding author. \textit{Email address:} \textbf{gorkavol@gmail.com} (Volodymyr Gorkavenko)}, O.V.~Barabash${}^1$,
  T.V.~Gorkavenko${}^1$,  
O.M.~Teslyk${}^1$,\\
 A.O. Zaporozhchenko${}^1$, Junji Jia${}^3$,  A.I. Yakimenko${}^{1,4,5}$, E.V.~Gorbar${}^{1,2}$\\
${}^1$ \it \small Faculty of Physics, Taras Shevchenko National University of Kyiv,\\
\it \small 64, Volodymyrs'ka str., Kyiv 01601, Ukraine\\
${}^2$ \it \small Bogolyubov Institute for Theoretical Physics, National Academy of Sciences of Ukraine,\\
\it \small 14-b Metrolohichna str., Kyiv 03143, Ukraine\\
${}^3$ \it \small School of Physics and Technology, Wuhan University, 299 Bayi Road, Wuhan, Hubei Prov., China\\
${}^4$ \it \small Dipartimento di Fisica e Astronomia ’Galileo Galilei’, Universit'a di Padova,\\ \it \small via Marzolo 8, 35131 Padova, Italy\\
${}^5$ \it \small Istituto Nazionale di Fisica Nucleare, Sezione di Padova, via Marzolo 8, 35131 Padova, Italy
}
\date{}

\maketitle
\setcounter{equation}{0}
\setcounter{page}{1}%

\begin{abstract}
Dynamical friction and stellar orbital motion in spiral galaxies with dark matter composed of ultralight bosons in the state of rotating Bose-Einstein condensate (BEC) are studied. It is found that the dynamical friction force is significantly affected by the topological charge of the vortex structure of the BEC core with the strongest effect at distances near the galactic center. It is also shown that the ultralight dark matter self-interaction plays an important role in studying the dynamical friction. 

Keywords: {dark matter, ultralight dark matter bosons, vortex structure}
\end{abstract}

\section{Introduction}

The problem of nature and composition of dark matter (DM) is one of the most pressing problems of modern physics and astrophysics. DM makes up 26.8\% of the mass-energy composition of the Universe \cite{Bertone:2004pz} and is assumed to consist of beyond the Standard Model particles which do not interact with Standard Model particles or interact with them very feebly. For a recent review of observational evidence for the distribution of dark matter in galaxies, see \cite{Salucci:2018hqu}. There are many candidates for the explanation of DM including primordial black holes \cite{Khlopov:1985fch,Carr:2016drx}.
Among popular candidates for DM particles are weakly interacting massive particles (WIMPs) with mass of $\sim$100 GeV \cite{Arcadi:2017kky}, sterile right-handed neutrinos with masses of several keV  \cite{Asaka:2005,Asaka_2:2005}, and axion-like (pseudoscalar) particles with masses of $\mu$eV \cite{Adams:2022pbo}.   At the moment, none of these particles has been directly observed and the corresponding models are known as the cold dark matter (CDM) models \cite{Bertone}.

In this paper, we will consider another attractive possibility of ultralight DM composed of spinless bosons with masses $\sim 10^{-23}-10^{-21}\,\mbox{eV}$ whose enormous occupation number results in the formation of the Bose-Einstein Condensate (BEC) with superfluid properties   { (for a review and discussion of the observable restrictions on the bosonic particle mass, see, e.g., \cite{Ferreira,Calmet:2015fua,Matos:2023usa})}. Ultralight dark matter (ULDM) models have very interesting phenomenology. Because of the very large de Broglie wavelength of order kpc, the formation of small-scale
structures in ULDM models is suppressed, unlike CDM models which predict too many dwarf galaxies and too much dark matter in the innermost regions of galaxies.

The fascinating physics of superfluid DM leads to a number of interesting ULDM structures whose observational signatures might help to elucidate the basic properties of DM particles.
Recently, it was found that wavelike ULDM may resolve lensing anomalies of Einstein rings in gravitationally lensed images \cite{Schive:2023}. Further, quantum wave interference of CDM without self-interaction (also known as fuzzy DM \cite{Hu:2000ke},  {ultralight pseudo Nambu-Goldstone boson appearing in the late time cosmological phase transition theories was proposed in \cite{Sin:1992bg} as a natural realization of fuzzy dark matter}) produces stochastically fluctuating density granulation which could drive significant disc thickening, providing a natural explanation for galactic thick discs \cite{Ostriker:2024}.  {It was shown that repulsive self-interactions, even weak, could be important for galactic halo composed of ultralight bosons \cite{Lee}. A scalar field model of dark matter in spiral galaxies was proposed in \cite{Matos} which produces the stellar orbital velocity profile in good agreement with
observations. For purposes of this work, it is important that ULDM could form stable vortex structures \cite {Nikolaieva:2021owc} which survive even after head-on collision \cite{PhysRevD.108.023503}}.
Repulsive self-interaction is critical for the existence of stable vortex solutions.

Given that direct observation of DM remains elusive, we focus on the influence of dynamic friction on baryonic matter. 
We consider how different soliton structures of bosonic ULDM affect the motion of stars in galaxies.
We concentrate on stellar motion because it admits semi-analytic treatment of dynamical friction, unlike the case of motion of supermassive black holes in the ULDM environment near the galactic centers. Indeed, a recent analysis in \cite{Boey:2024dks} showed that the presence of a moving supermassive black hole significantly affects  ULDM soliton, leading to dynamical friction oscillations, requiring numerical analysis. Further, it was shown that stochastic density fluctuations of ULDM may stall the inspiral of supermassive black holes or globular clusters due to dynamical friction at radii of a few hundred parsecs and can heat and expand the central regions of galaxies \cite{Tremaine}.

The classical work by Chandrasekhar \cite{Chandrasekhar} on dynamical friction experienced by a moving star due to the fluctuating gravitational force of neighbouring stars was generalized to the case of the motion of stars in ULDM in the absence of self-interaction  {(the case of fuzzy DM \cite{Hu:2000ke})} in Ref.\cite{Hui:2016ltb} where analytic expression for dynamical friction was derived. Later the obtained results were numerically checked in \cite{Lancaster_2020,Wang}.  {The ULDM self-interaction} modifies the dynamical friction force and its role for the star's motion in self-interacting  {ultralight dark matter} was studied in \cite{Glennon}.

The ground and vortex states of the ULDM yield different DM density and velocity distributions. Therefore, the doughnut-like density distribution (vanishing at the vortex core) and vortex flows (rapidly increasing at the vortex axis) of the BEC core of vortex soliton should directly affect the dynamical friction and, hence, star motion and their relaxation time.
Since stellar motion is directly observed, the problem of the impact of the vortex structure of ULDM on the dynamical friction provides the main motivation for the study performed here, which to the best of our knowledge was not considered in the literature.

The paper is organized as follows. Dynamical friction acting on moving stars in ULDM with trivial and nontrivial vortex structures is considered in a simplified approach in the absence of the ULDM self-interaction in Sec.\ref{sec:dynamical}. Taking the ULDM self-interaction into account, the stellar orbital motion is studied in Sec.\ref{sec:self-interaction}. Conclusions are drawn in Sec.\ref{sec:Conclusion}.

\section{Model and dynamical friction in simplified model}
\label{sec:dynamical}

When a star moves in dark matter in the form of the BEC, it produces perturbation in galactic BEC leading to an overdense wake. This wake induces a drag force on a moving star resulting in dynamical friction. Conceptually, the situation is quite similar to the motion of bodies in fluids.
Since the de Broglie wavelength of the BEC is much larger than stellar radii, we could approximate stars as point objects. As stars move through the galactic BEC, a gravitational wake forms behind them and the
associated overdensity causes resistance producing dynamic friction
acting on moving stars.

Since we are interested in the impact of vortex states of ULDM on the motion of stars and the repulsive self-interaction of ULDM is crucial for the existence of stable  {vortex} structures, we employ in our analysis of dynamical friction the analytic results obtained in \cite{Glennon,Berezhiani:2023vlo}, which determined the coefficient of dynamical friction in self-interacting ULDM.
 
Let us begin our analysis by reminding the system of the Gross-Pitaevskii-Poisson equations  {\cite{Ferreira:2020fam,Calmet:2015fua}} which govern the dynamical evolution of self-gravitating and self-inter\-act\-ing BEC field $\psi$ and the gravitational potential $\Phi$ 
\begin{equation} 
i\hbar\frac{\partial\psi}{\partial t} = \left(-\frac{\hbar^{2}}{2m}\nabla^{2} + gN|\psi|^{2} + m\Phi_{g} \right)\psi,
  \label{GPP 1}
\end{equation}
\begin{equation}
  \nabla^{2}\Phi_{g} = 4\pi GmN|\psi|^{2},
  \label{GPP 2}
\end{equation}
where $g = {4 \pi \hbar^{2}a_{s}}/{m}$ is the coupling strength of the self-interaction of DM particles with mass $m$, $a_{s}$ is the $s$-wave scattering length, $N$ is the number of boson particles, $\hbar$ is the Planck constant, and $G$ is the gravitational constant.  {The first term on the right-hand side of the Gross-Pitaevskii equation (\ref{GPP 1}) corresponds to the kinetic energy. Repulsive self-interaction is described by the nonlinear term in Eq. (\ref{GPP 1}) with $g>0$.  The last term contains the self-consistent gravitational potential $\Phi$ which supports the formation of a localized BEC core. } 

As for the DM density in the BEC state, it was investigated for the case of the Milky Way galaxy in Refs.\cite{Bryan:1997dn,Schive:2014hza,Lin:2019yux,Pozo:2023xfh}.
 {Approximate variational and numerical solutions to the GPP equations were found in \cite{Nikolaieva:2021owc,Korshynska:2023kxa}. In present work, we use a variational approach for stationary states with the following ansatz for $\psi$ in cylindrical coordinates $z,r,\phi$: 
\begin{equation}
\psi(t, r, \phi, z) = A \left(\frac{r}{R}\right)^{s}e^{-i\mu t -\frac{r^{2}}{2R^{2}} - \frac{z^{2}}{2(R\eta)^{2}} + is\phi}
\label{variational ansatz},
\end{equation}
where $R$ and $\eta$ are variational parameters, $A$ is the normalization constant, and $\mu$ is the chemical potential}. The quantum number $s$ is the
topological charge of the galactic BEC condensate. While the case $s=0$ corresponds to the ground BEC state, the $s=1$ state describes the BEC soliton rotating around the center of the galaxy. States with higher topological charge $s>1$ are unstable \cite{Dmitriev:2021utv,Nikolaieva:2021owc}.  {The gravitational potential $\Phi$ can be found by solving the Poisson equation (\ref{GPP 2}). This equation allows for an analytical solution in the $s=0$ \cite{Korshynska:2023kxa} state and $\Phi$ can be determined numerically for $s>0$.}

\begin{figure}[t!]
  \begin{center}
    \includegraphics[width=0.7\textwidth]{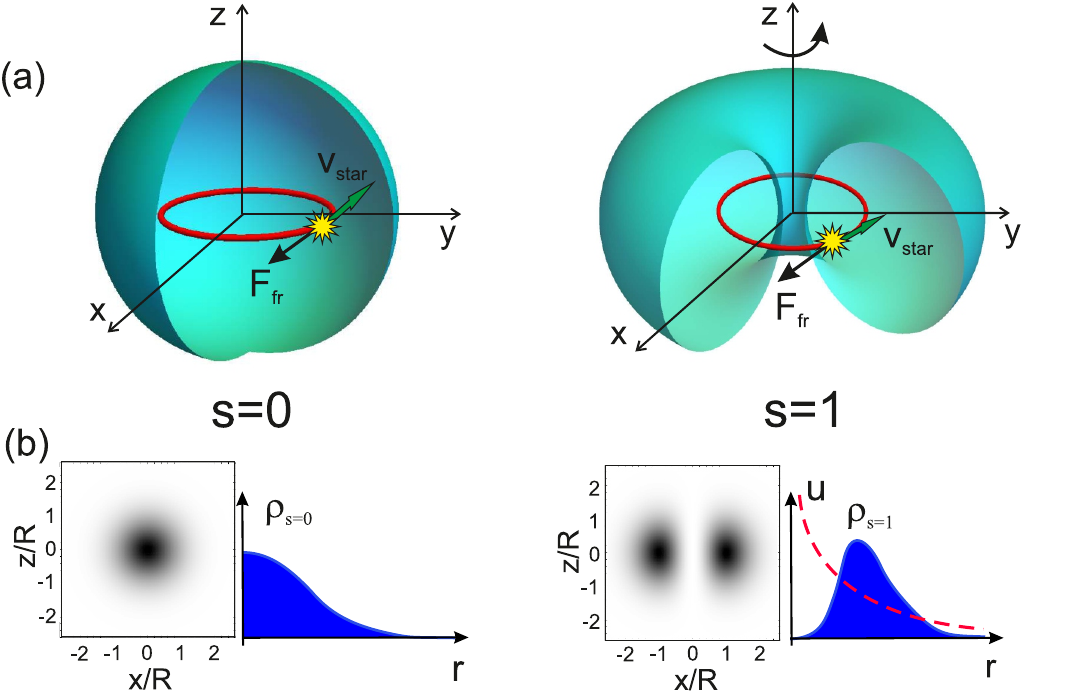}
    \caption{Schematics of a stellar object in a circular orbit subject to effective dynamical friction caused by  {ULDM}. (a) The left panel displays the density isosurface of  {ULDM} solitonic core with zero angular momentum (state $s=0$). The right panel depicts the Bose-Einstein condensate (BEC) core in the vortex state ($s=1$, the direction of circulation  {is} shown by arrow). The dynamical friction force acting on a moving star, $F_{fr}$, is shown by arrow. (b) Schematics of the condensate density distribution  {(shaded blue for the radial distribution and gray-scale density plot for the distribution in $(x,z)$-plane)} as a function of the radial coordinate $r$ for the soliton (left panel) and vortex (right panel) states. The red dashed line shows the radial dependence of the ULDM velocity $u$ in the vortex state.}
     \label{fig_full_density}
    \end{center} 
\end{figure}

It is very important for the study of dynamical friction that vortex states of BEC affect the DM profile differently. Here we use the DM profile $\rho_{DM}\sim |\psi|^2$ obtained in the case of the Milky Way galaxy in \cite{Korshynska:2023kxa} for distances from the galaxy center less than 1 kpc for the trivial ground $s=0$ and nontrivial $s=1$ vortex states given by (for $r > 1$ kpc, the Navarro-Frenk-White profile \cite{Navarro:1995iw} was used)
\begin{equation}\label{rhoDMs0}
    \rho_{s=0}=0.34\cdot 10^{-17} e^{-\left(\frac{r_{sph}}{0.61}\right)^2} \frac{\rm kg}{\rm m^3},\quad r<1\, kpc
\end{equation}
\begin{equation}\label{rhoDMs1}
    \rho_{s=1}=0.88\cdot 10^{-17} \left(\frac{r}{0.4}\right)^2 e^{-\left(\frac{r}{0.4}\right)^2 - \left(\frac{z}{0.58}\right)^2} \frac{\rm kg}{\rm m^3},\quad r< 1\, kpc,
\end{equation}
where $r_{sph}^2=r^2+z^2$ is the radial spherical distance and $r$, $z$ are coordinates in the cylindrical coordinate system measured in kpc, see Fig.\ref{fig_full_density}. This density function was obtained for
the Milky Way halo mass $M= 1.3 \times 10^{12}M_{\odot}$,  {where $M_{\odot}$ is the solar mass}, and radius $R_{\mathrm{halo}} = 287\, kpc$.
We use the model described in \cite{Chavanis:2018pkx} and studied in \cite{Korshynska:2023kxa} and choose the
DM particle mass $m = 2.92 \times 10^{-22}\, eV/c^{2} = 0.52 \times 10^{-57}\, kg$  {with the} scattering length $a_{\mathrm{s}} = 8.17 \times 10^{-77}$ meter.

For the vortex state $s=1$, the velocity of DM is given by \cite{Korshynska:2023kxa}
\begin{equation}\label{DMspeed}
    \textbf{u}=c\frac{r_u}{r}\textbf{e}_\phi,
\end{equation}
where $r_u=2.2\cdot 10^{-5}$ kpc. To determine dynamical friction, we use the rotation curve of the Milky Way galaxy  {observed by astrophysical means} \cite{Lambourne:2015cfz,Sofue:2008wt,Sofue:2023}  {in the interval 10 pc $<r<10^3$ pc and discussed in detail in \cite{Sofue:2023}. The first reason for considering distances larger 10 pc is connected with the presence of the supermassive black hole Sgr A* in the Milky Way center, which could severely affect the DM density profile for $r \leq 10$ pc producing a dark matter spike with power law profile \cite{Nampalliwar}. 
Another reason is that near the center of the galaxy, the contribution to the matter density from the baryonic component is significant \cite{Jusufi:2020cpn}. 
Therefore, we consider larger distances where the impact of the supermassive black hole and baryonic component is not essential.
We restrict our consideration to distances less 1 kpc where bosonic DM is in the BEC state. For larger distances, there is isothermal ULDM halo \cite{Chavanis:2018pkx} where DM density is no longer given by Eqs.(\ref{rhoDMs0}) and (\ref{rhoDMs1}).}

In this study, we quantify the effect of dynamic friction on moving stars. Since the self-interaction of ULDM particles is crucial for the existence of stable vortex states of BEC, we leave
the analysis of dynamical friction of the $s=1$ state for the next section. Here we
will consider dynamical friction only for the ground $s=0$ state which can be realized also in the fuzzy DM.

Although the DM profile of the $s=0$ state in Eq.(\ref{rhoDMs0}) was found in \cite{Korshynska:2023kxa} in a model with ULDM self-interaction, we will assume that this profile is valid also in the case of fuzzy DM to compare the obtained results for dynamical friction with those found in the correct approach in the next section where self-interaction is taken into account.

In the case of fuzzy DM, dynamical friction was studied in \cite {Hui:2016ltb}. For parameter $\beta\ll 1$, where
$$
\beta=\frac{2\pi G m_{star} m}{\hbar v_{star}},
$$
the friction force equals \cite{Hui:2016ltb}
\begin{equation}\label{WittenForce}
   F_{fr}=4\pi \frac{G^2 m_{star}^2 \rho_{DM}}{v_{star}^2} C(\Lambda),
\end{equation}
where
\begin{equation}
   \Lambda= \frac{kr}{\beta}=\frac{v^2_{star}r}{G m_{star}}  
\end{equation}
is the gravitational counterpart of the Coulomb logarithm\footnote[1]{ {
In plasma kinetic theory the Coulomb logarithm appears in the description of charged particle collisions with the long-range Coulomb interaction (clearly, a similar situation is realized in the case of the long-range Newtonian potential).
The corresponding logarithmic divergence at large $r$ can be eliminated by introducing a cut-off at small scattering angle. Physically, the elimination of divergences is explained by the fact that the Coulomb field of the charged particle at large distances is shielded by other charged particles \cite{coulomb,pitaevskii2012physical}.} }, $m_{star}$ is the star's mass, $v_{star}$ is the absolute value of the star's velocity, $k=mv_{star}/\hbar$, and $r$ is the radius of the stellar orbit.
Since $m_{star}$ is much less than the host system in which it resides, we have the following simple relation for the function $C$ \cite{Hui:2016ltb}:
\begin{equation}\label{Crelat1}
   C(\Lambda) =
      \frac{1+\Lambda}{\Lambda} \mbox{arctanh}\frac{\sqrt{\Lambda^2 + 2\Lambda}}{1 + \Lambda} - \sqrt{
 1 + \frac{2}{\Lambda}}.
\end{equation}
 {In our case, parameter $\Lambda \gg 1$ for distances 10 pc $<r<10^3$ pc and it is convenient to use more simple asymptotic expression for the function $C$ given by 
\begin{equation}\label{Crelat1a}
 C(\Lambda) =\ln(2 \Lambda) (1 + 1/\Lambda) - 1.
\end{equation}
}
 Equation \eqref{WittenForce} implies that the friction force increases for stars with small velocities and in the region with  {high DM density}, i.e., near the galactic center.

Our main quantity of interest is a characteristic time when a moving star changes notably its velocity \cite{Hui:2016ltb}
\begin{equation}
T =\frac{v_{star}m_{star}}{F_{fr}}.
\label{eq:T10-time}
\end{equation}
Using Eq.(\ref{WittenForce}), we find
\begin{equation}\label{eq:T10}
    T=\frac{v^3_{star}}{G^2 m_{star} \rho_{DM}}\frac{1}{4\pi C(\Lambda)}.
\end{equation}

Our results for $T$ in ULDM without self-interaction are presented by dash-dotted line in Fig.\ref{fig:timekhuri}, where we used $m_{star}=M_{\odot}$.  As one can see, $T$ monotonously grows with $r$.  {The reason is that while the DM density is almost constant at small distances 0.1 kpc $<r<$ 0.34 kpc, $v_{star}$ increases linearly. Then at larger distances $r>0.34$ kpc stellar velocity $v_{star}$ does not change much but DM density decreases notably}. 
Hence $T$ should grow in view of Eq.(\ref{eq:T10}). In the next subsection, we will compare these results for the $s=0$ state with those obtained in a more accurate approach where the ULDM self-interaction is taken into account.

\section{Motion of stars in ULDM with vortex structure}
\label{sec:self-interaction}

To determine dynamical friction of stars moving in ULDM with different vortex structures, we assume that stars move on circular orbits with constant velocity and constant angular velocity $\Omega$ in the self-interacting ULDM and use the expression for dynamical friction derived in Ref.\cite{Berezhiani:2023vlo}.

As we mentioned in the Introduction, a stable DM structure for the case of rotating BEC with the quantum number $s=1$ is possible only for self-interacting ULDM particles  {with coupling constant of repulsive self-interaction exceeding some} critical value which is fixed by the soliton core's observable radius and the DM particle mass \cite{Nikolaieva:2021owc,Calmet:2015fua}. Following Ref.\cite{Korshynska:2023kxa}, we take $a_{s}=8.17 \cdot 10^{-77}$ meter and $m=0.52 \cdot 10^{-57}$ kg. Then we find $g/\hbar^2=1.97\cdot 10^{-18}$ meter/kg. 

For our purposes, it is sufficient to consider dynamical friction in the direction tangential to a circular orbit.
 {The dynamical friction force $\vec F_{fr}=-m_{star} \vec \nabla \phi(\vec r,t)$ is determined by the gravitational potential $\phi$ induced by perturbation of the DM density by a moving body, for example, a star with mass $m_{star}$. It can be found by solving the system of the Euler and continuity equations for the DM density and velocity as well as taking into account the Poisson equation for the gravitational potential (see, for details \cite{Hui:2016ltb,Desjacques_2022}).} 
For a star with mass $m_{star}$ moving at constant velocity $v_{star}$ on a circular orbit of radius~$r$, the tangential dynamical friction equals \cite{Desjacques_2022,Buehler:2022tmr,Berezhiani:2023vlo} 
\begin{equation}\label{forcekhuriFull}
     F_{fr}=\frac{4\pi G^2 m_{star}^2 \rho_{DM}}{ {c_s^2}}\mathcal{F},
\end{equation}
where function $\mathcal{F}$ is given in the form of the sum over the magnetic quantum number $m_l$ and the azimuthal quantum number $\ell$ (note that the derivation of this dynamical friction force is quite similar to that given by Ostriker's formula \cite{Ostriker} for the motion of a perturber in gaseous medium)
\begin{equation}
    \mathcal{F}= \sum_{\ell=1}^{\ell_\text{\tiny max}}\sum_{m_l=-\ell}^{\ell-2}\gamma_{\ell m_l}\text{Im}\left(S_{\ell,\ell-1}^{m_l}-{S^{m_l+1}_{\ell,\ell-1}}^*\right),
\label{FDF1}
\end{equation}
\begin{multline}
    \gamma_{\ell m_l}=  (-1)^{m_l} \frac{(\ell-m_l)!}{(\ell-m_l-2)!}\\ \times\left\{{\Gamma\left(\frac{1-\ell-m_l}{2}\right)\Gamma\left(1+\frac{\ell-m_l}{2}\right)\Gamma\left(\frac{3-\ell+m_l}{2}\right)\Gamma\left(1+\frac{\ell+m_l}{2}\right)}\right\}^{-1}\,,
\end{multline}
\begin{align}
    S_{\ell,\ell-1}^{m_l}&= \frac{\pi {\rm i}}{2 \sqrt{1+\frac{m^2_l}{\ell_{\rm q}^2}} }\biggr[(-1)^{1+\theta(m_l)}j_\ell\big(\ell_{\rm q} \mathcal{M}  f^-_{m_l} \big)j_{\ell-1}\big(\ell_{\rm q} \mathcal{M}  f^-_{m_l} \big)
    \nonumber \\
    & 
    ~~~~~~~~~~~~~~~ ~+{\rm i} j_\ell\big( \ell_{\rm q} \mathcal{M}  f^-_{m_l} \big)y_{\ell-1}\big(\ell_{\rm q} \mathcal{M}  f^-_{m_l} \big) -j_\ell\big( {\rm i} \ell_{\rm q} \mathcal{M}  f^+_{m_l} \big)h_{\ell-1}^{(1)}\big( {\rm i} \ell_{\rm q} \mathcal{M}  f^+_{m_l} \big)\biggr]\,,\label{Slm}
\end{align}
$j_l(x)$ and  {$y_l(x)$} are the  spherical Bessel functions of the first and second kind,  $h_l^{(1)}(x)$ is the spherical Hankel function of the first kind, $\theta(x)$ is the Heaviside step-function, and
\begin{equation}
f^{\pm}_{m_l} = \sqrt{2}\left(\sqrt{1+\left({m_l}/{\ell_{\rm q}}\right)^2}\pm1\right)^{1/2},
\label{fpm}
\end{equation}
where parameter $\ell_q$ is defined as
\begin{equation}\label{Lq}
   \ell_q=\frac{rm c_s}{\hbar \mathcal{M}}.
\end{equation}

The important characteristic of the problem of dynamical friction is the Mach number
\begin{equation}\label{Mach}
    \mathcal{M}=\frac{v}{c_s},
\end{equation}
where $v$ is the absolute value of the relative velocity of the star and ULDM velocity $\mathbf{u}$ and  {$c_s$ is the adiabatic sound speed} of DM superfluid. The latter is given by the first derivative of pressure concerning density \cite{Ferreira:2020fam}
\begin{equation}\label{soundveloc}
 c_s=\frac{\sqrt{g\rho_{DM}}}{m}.
\end{equation}
Note that pressure is proportional to the coupling constant of self-interaction $g$, therefore,  {$c_s$} vanishes in fuzzy dark matter. 

\begin{figure}[t]
    \centering
    \includegraphics[width=0.5\textwidth]{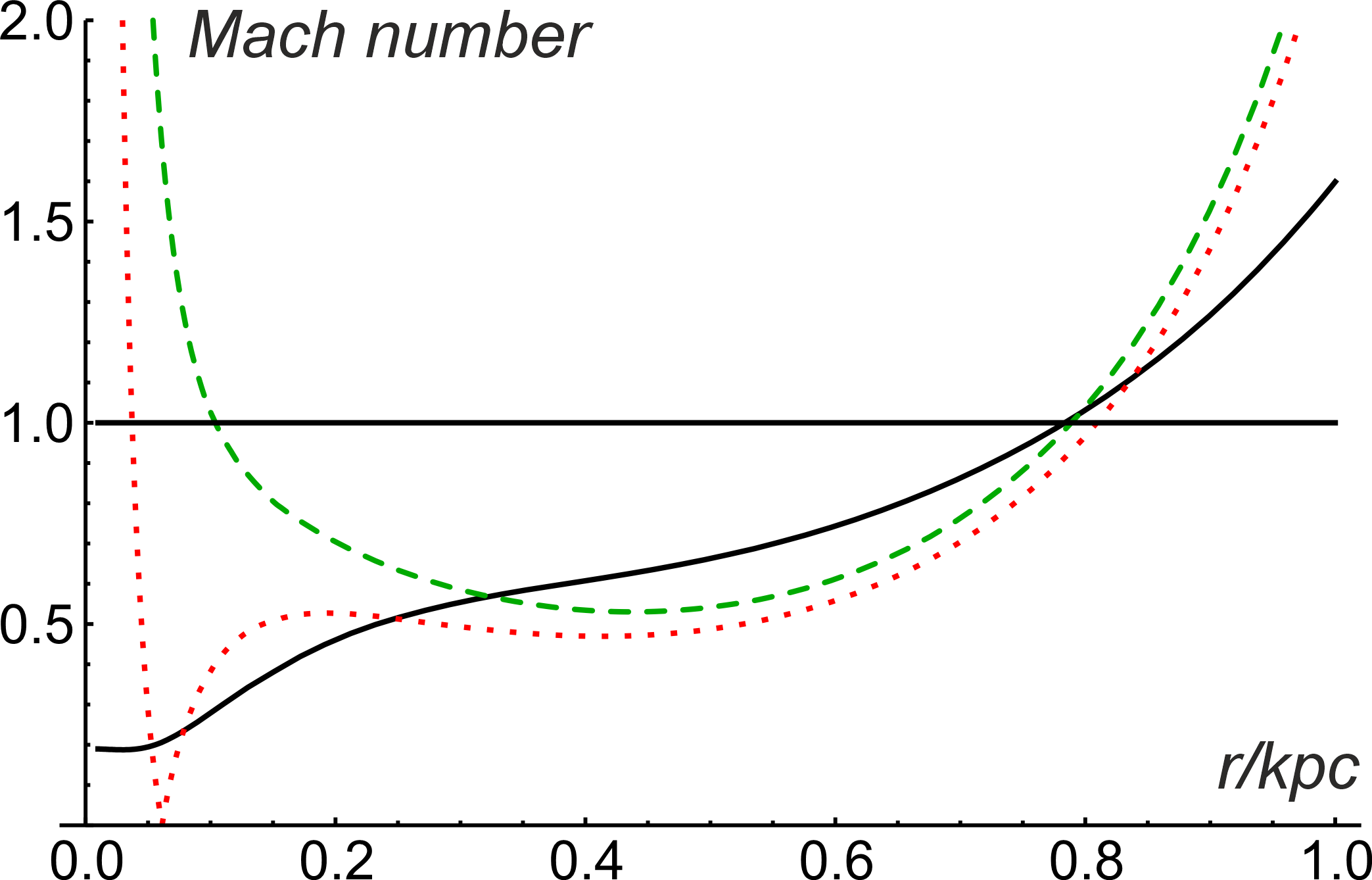}
    \caption{The Mach number as a function of distance $r$ to the galactic center for different vortex states of  {ULDM}. 
We consider the ground BEC state $s=0$ (solid black line) and vortex BEC state $s=1$  for the cases where a probe star corotates (dotted red line) and counter-rotates (dashed green line) with respect to DM superfluid. The horizontal black solid line corresponds to the Mach number $\mathcal{M}=1$.}
    \label{fig:Mach}
\end{figure}

For the ground BEC state $s=0$, where $\mathbf{u}=0$, the relative velocity $\mathbf{v}$ coincides with the star's velocity $v=v_{star}$. In the vortex state $s=1$, there are two possiblities of counter-rotating ($v=v_{star}+u$) and corotating star motion ($v=|v_{star}-u|$). Since the BEC velocity rapidly decreases with distance from the galactic center, the difference between these two cases is significant only at a small distance from the galactic center. The Mach number as a function of distance to the  {galactic center} $r$ is plotted in Fig.\ref{fig:Mach} for the ground ($s=0$) and vortex ($s=1$) for the cases where a probe star corotates and counter-rotates with respect to DM superfluid.

\begin{figure}[h]
    \centering
    \includegraphics[width=0.5\textwidth]{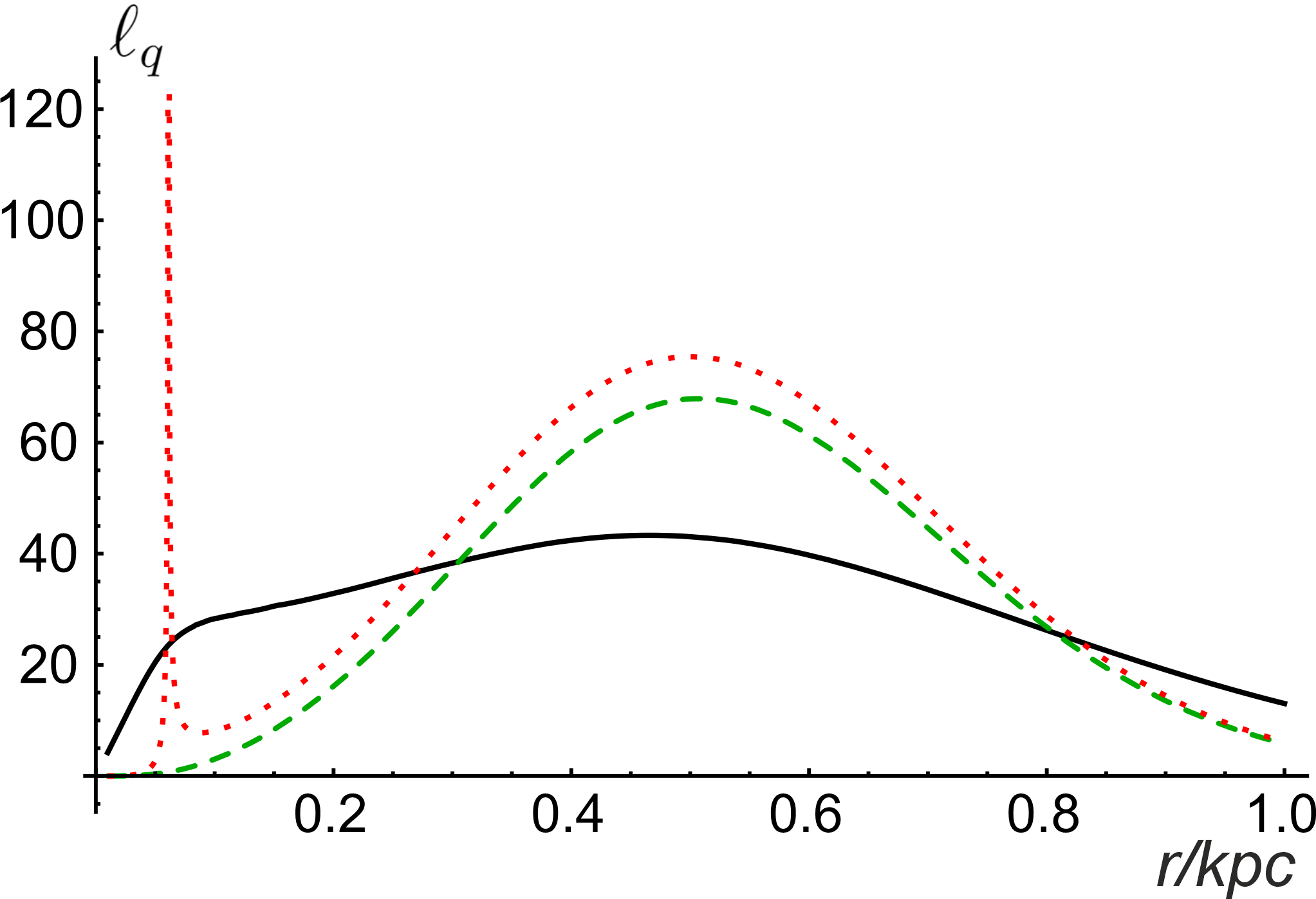}
    \caption{Parameter $\ell_q$ as a function of distance $r$ to the galactic center for the  $s=0$ ground BEC state (solid black line) and the cases of counter-rotating (dashed green line) and corotating (dotted red line) star's orbits for the $s=1$ vortex BEC state.}
    \label{fig:lq-1}
\end{figure}

Let us discuss now parameter $\ell_q$. This parameter as a function of distance $r$ to the center of the galaxy is presented in Fig.\ref{fig:lq-1} for the ground $s=0$ BEC state and $s=1$ vortex BEC state for corotating and counter-rotating stellar orbits. As one can see, parameter $\ell_q$ is typically much larger than the unity.
Further, the maximum azimuthal quantum number $\ell_{max}$ up to which the sum over $l$ is performed is determined by the size~$R_\text{\tiny probe}$ of probe  {body} and orbit's radius~$r$
\begin{equation}
\ell_\text{\tiny max} =\frac{\pi r}{R_\text{\tiny probe}}\,.
\label{lmax}
\end{equation}
Since the radius of a star is usually much less than its orbit's radius, we have $\ell_{max} \gg 1$. Since series \eqref{FDF1} is convergent, we could introduce instead of $\ell_{max}$ a cut-off whose value is chosen by the required accuracy of calculations. 

  It is worth mentioning that there is a simple analytic expression which approximates the tangential component of the friction force for $\mathcal{M}<1$ and $\ell_q\rightarrow \infty$ \cite{Berezhiani:2023vlo}
\begin{equation}\label{forcekhuri}
    F_{fr}=\frac{4\pi G^2 m_{star} \rho_{DM}}{c_s^2}\frac{{\rm arctanh} \mathcal{M}-\mathcal{M}}{M^2}.
\end{equation}
Comparing this relation with \eqref{forcekhuriFull} one can see that function $\mathcal{F}$ in this case has a simple form
\begin{equation}\label{Fappth}
     \mathcal{F}=\frac{{\rm arctanh} \mathcal{M}-\mathcal{M}}{M^2}.
\end{equation}
As one can see from Fig.\ref{fig:Mach}, there is a region where the Mach number is less the unity, i.e., $\mathcal{M}<1$,  but parameter $\ell_q$, in view of Fig.\ref{fig:lq-1}, is large in this region but not infinite. 
 This interval defines the region where the above approximate formula for the dynamical friction force could be applied.

\begin{figure}[h]
    \centering
    \includegraphics[width=\linewidth]{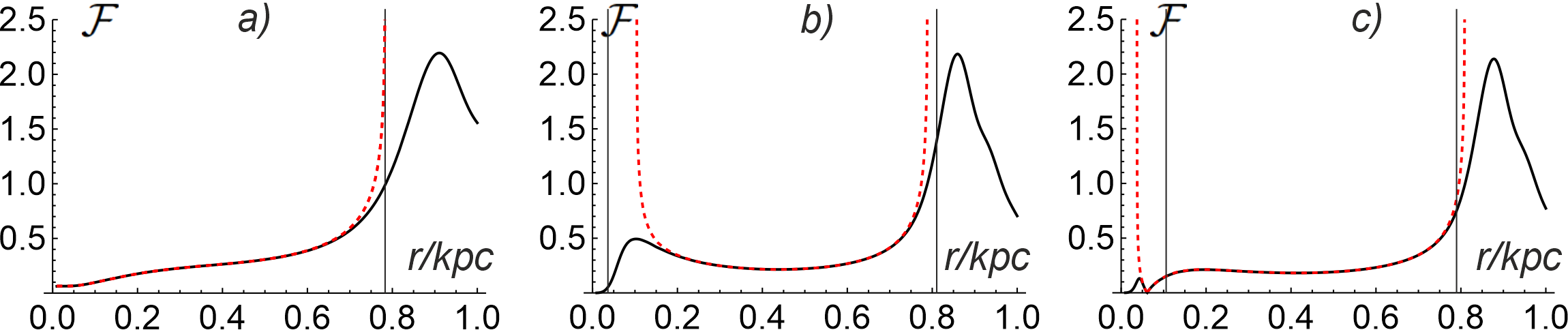}
    \caption{ {The function $\mathcal{F}$ as a function of distance $r$ to the galactic center for the $s=0$ ground BEC state (panel \textit{a}) and  corotating (panel \textit{b}) and counter-rotating (panel \textit{c}) stellar orbits for the $s=1$ vortex BEC state. Solid black lines correspond to the function $\mathcal{F}$ given by Eq.\eqref{FDF1} and dashed red lines correspond to $\mathcal{F}$ given by the approximate formula \eqref{Fappth}. Two vertical solid lines restrict the region where $\mathcal{M}<1$ and the approximate analytic formula (\ref{Fappth}) can be applied}.}
    \label{fig:Fkry}
\end{figure}

 {For stars in their rotation motion in the Milky Way, our results for function $\mathcal{F}$ defined by Eq.\eqref{FDF1} and the approximate analytic formula \eqref{Fappth} are given in Fig.\ref{fig:Fkry}. It can be seen that the results agree quite well for the region where the Mach number is $\mathcal{M}\lesssim1$, 
otherwise, the friction force defined by the approximate expression \eqref{forcekhuri} blows up.}

\begin{figure}[h]
    \centering
    \includegraphics[width=0.5\textwidth]{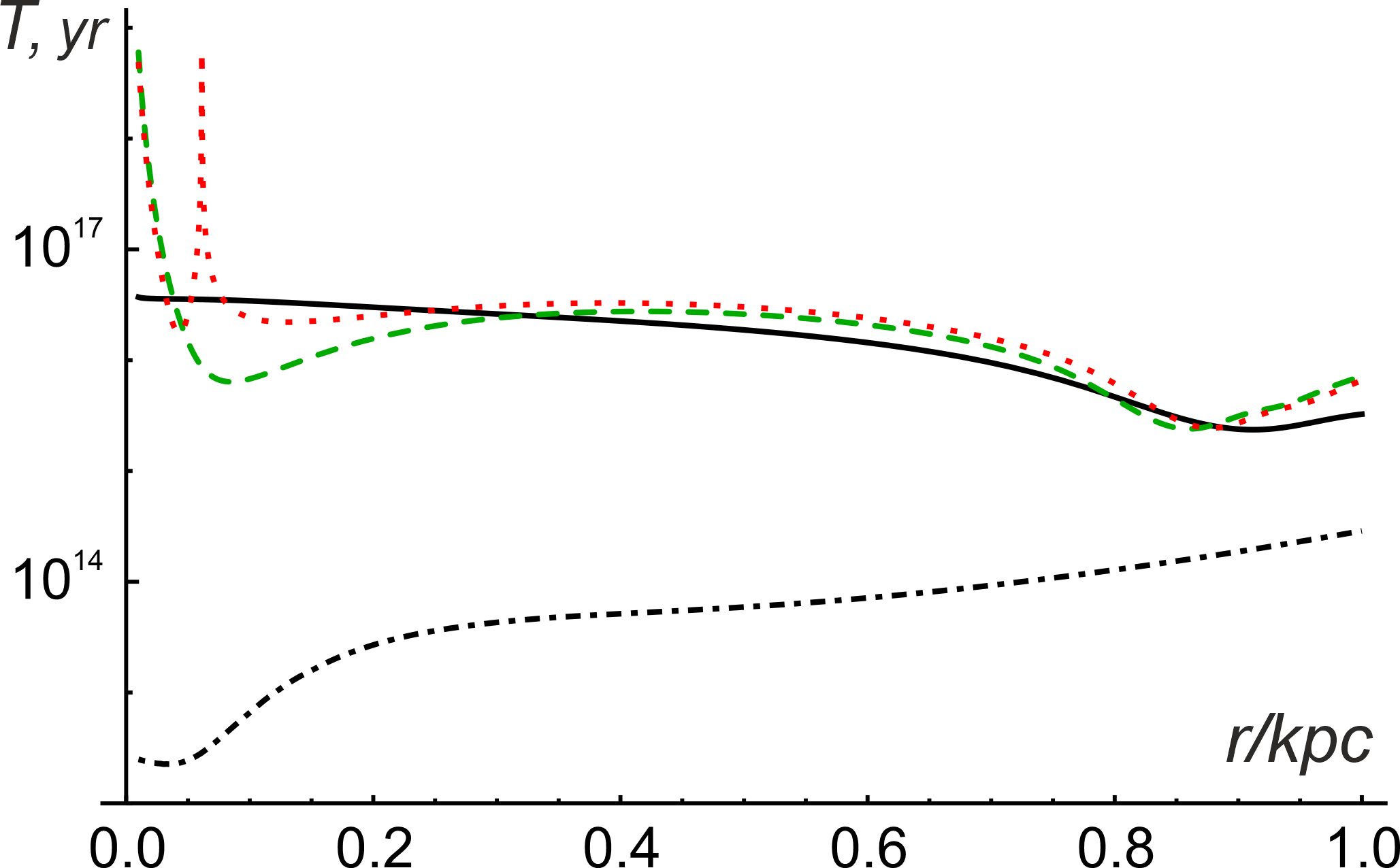}
    \caption{The characteristic time  {in years} needed to decrease the stellar velocity in self-interacting DM as a function of distance to the galactic center for the DM particle mass $m=2.92 \cdot 10^{-22}\,eV/c^2$ for the  $s=0$ ground BEC state (solid black line) and the cases of counter-rotating (dashed green line) and corotating (dotted red line) stellar orbits for the $s=1$ vortex BEC state. The dash-dotted black line corresponds to fuzzy DM without self-interaction plotted using Eq.\eqref{eq:T10}.}
    \label{fig:timekhuri}
\end{figure}

Having defined the dynamical friction force for ULDM with self-interaction, we can easily determine the characteristic time for a star to decrease its velocity using Eq.\eqref{eq:T10-time}. The corresponding results are plotted by solid, dashed, and dotted lines in Fig.\ref{fig:timekhuri} for distances $r> 1$ pc in the $z=0$ plane and  {and can be summarized as follows}. (i) The characteristic time needed for a star to decrease its velocity is practically the same for the ground $s=0$ and vortex $s=1$  {states} for $r\gtrsim 0.3$ kpc, where the ULDM velocity $\mathbf{u}$ is small and can be neglected. (ii) For  {$r\lesssim 50$ pc}, $T$ is approximately the same for the counter-rotating and corotating motion in the $s=1$ state for which the star's velocity can be neglected compared to the ULDM velocity at small $r$. Clearly, the decrease of ULDM density in the $s=1$ state compared to the $s=0$ state, see Fig.\ref{fig_full_density}, strongly diminishes the friction force and makes $T$ very large at a small distance. (iii) $T$ strongly increases for the co-rotating case in the $s=1$ state at $r \approx 0.06 $ kpc, where the relative velocity of the star's velocity and the ULDM velocity vanishes. This means  {that} the ULDM rotation makes stellar orbital motion more stable at a certain distance from the galactic center.

Finally, let us compare the results for the ground $s=0$ state found in the simplified approach using Eq.(\ref{WittenForce}) for the dynamical friction force where the ULDM self-interaction was not taken into account (dash-dotted black line in Fig.\ref{fig:timekhuri}) with the results obtained  {in the case where the ULDM self-interaction is} taken into account (solid black line). Clearly, we see that the ULDM self-interaction is very important for the relaxation time.

\section{Conclusions}
\label{sec:Conclusion}

For  {ULDM in the state of BEC, we considered} two different configurations of BEC: ground (not rotating) and vortex states. We studied the dynamical friction force acting on stars moving in ultralight dark matter with different soliton vortex structures and investigated the characteristic time $T$ for a star to decrease its velocity. The main results of our investigation are presented in Fig.\ref{fig:timekhuri},  {where the characteristic relaxation time for stars in their rotation in the Milky Way is displayed} and are shown by solid black ($s=0$ state), dashed green (the counter-rotating case of $s=1$ vortex DM state) and dotted red (the corotating case of $s=1$ vortex DM state) lines.  {According to this figure, the relaxation time always exceeds the Hubble time.}

We found that for large distances $r\gtrsim 0.3$ kpc, the  {characteristic} time needed to decrease the stellar velocity is practically the same for the ground $s=0$ and vortex $s=1$ states because the ULDM velocity is small. At small distances, $r\lesssim 50$ pc, $T$ is approximately the same for the counter-rotating and corotating  {stellar} motion in the $s=1$ state for which the star's velocity can be neglected compared to the ULDM velocity at small $r$.  $T$ strongly increases for the corotating case in the $s=1$ state at $r \approx 0.06$ kpc, where the relative velocity of the star's velocity and the ULDM velocity vanishes. This means the ULDM rotation makes the stellar orbital motion more stable at a certain distance from the galactic center. 

Overall, our analysis shows that the orbital motion of stars of the solar mass is stable in the Milky Way galaxy at least for distances from the galactic center   {$r>10$ pc} and the dynamic friction induced by BEC is not sufficient to affect strongly star's motion. Still, we found that the BEC structure and ULDM self-interaction are very important for the analysis of dynamical friction on star's motion in galaxies and much work remains to be done in this direction especially for objects much more massive compared to the Sun because the acceleration due to the friction force grows with mass. 

In addition, we confirmed the conclusion made in previous studies that the ULDM self-interaction essentially affects the dynamical friction. A similar conclusion was reached in recent studies \cite{Dewar,Fischer,Koo:2023gfm,Bromley:2023yfi} of the dynamical friction force for inspiraling black holes in the case of cold dark matter, where self-interaction may solve the final parsec problem of supermassive black holes merger.

\vspace{5mm}

\centerline{\bf Acknowledgements}
\vspace{5mm}

The authors thank K. Korshynska for useful discussions. V.M.G. is grateful to  Wuhan University for its hospitality during his stay there. A.I.Y. acknowledge support from the projects ‘Ultracold atoms in curved geometries’, 'Theoretical analysis of quantum atomic mixtures' of the
University of Padova, and from INFN.

\bibliographystyle{JHEP}
\bibliography{bibliography.bib}

\end{document}